# Can Maxwell's Fish Eye Lens Really Give Perfect Imaging? Part II. The case with passive drains

**Fei Sun[1] , Xiaochen Ge[1] and Sailing He[1,2*]**

[1] *Centre for Optical and Electromagnetic Research, JORCEP[KTH-LTU-ZJU Joint Research Center of Photonics], East Building #5, Zijingang campus, Zhejiang University (ZJU), Hangzhou 310058, China*

[2] *Department of Electromagnetic Engineering, School of Electrical Engineering, Royal Institute of Technology (KTH), S-100 44 Stockholm, Sweden*

*\* Corresponding author: sailing@kth.se*

**Abstract:** We use both FEM (finite element method) and FDTD (finite difference time domain method) to simulate the field distribution in Maxwell's fish eye lens with one or more passive drains around the image point. We use the same Maxwell's fish eye lens structure as the one used in recent microwave experiment [10]: Maxwell's fish eye lens bounded by PEC (perfect electric conductor) is inserted between two parallel PEC plates (as a waveguide structure). Our simulation results indicate that if one uses an active coaxial cable as the object and set an array of passive drains around the image region, what one obtains is not an image of the object but only multiple spots resembling the array of passive drains. The resolution of Maxwell's fish eye is finite even with such passive drains at the image locations. We also found that the subwavelength spot around the passive drain is due to the local field enhancement of the metal tip of the drain rather than the fish eye medium or the ability of the drain in extracting waves.

1.  Introduction

Maxwell's fish eye is a special optical imaging instrument with equal light paths [1-3]. Recently, Leonhardt claimed that a modified Maxwell's fish eye [bounded with a perfectly electrical conductor (PEC) boundary] can give perfect imaging [4, 5]. This has sparked some debate [6-9]. In our previous work [9], we have shown that the resolution of a modified Maxwell's fish eye lens without any drain is diffraction limited. In a recent experiment in the microwave region [10], Leonhardt and his co-workers used a planar waveguide structure for the 2D Maxwell's fish eye lens and set a passive coaxial cable at the image point to realize the drain in his earlier analytical model [4], where it was claimed that adding some drains at the image point in a Maxwell's fish eye lens can give a perfect image.

In this paper, we use both FEM (finite element method) and FDTD (Finite Difference Time Domain) to simulate in details the imaging performance of a Maxwell's fish eye lens of the waveguide structure (same as the one used in the recent microwave experiment [10]) with one or more passive drains around the image points. Our simulation results show that the subwavelength spot around the passive drain is due to the local field enhancement of the metal tip of the drain rather than the fish eye medium or the ability of the drain in extracting wave. It is therefore not meaningful to talk about image resolution for Maxwell's fish eye lens with such passive drains.

2.  Imaging performance in the presence of multiple drains

In this section, we use numerical methods (FEM and FDTD) to calculate the field distribution in a Maxwell's fish eye lens with one or more passive drains around the image point. The structure we use here is identical to that in the recent experiment [10]. The fish eye lens with PEC boundary (radius R=5cm) is inserted between two parallel PEC plates separated by 5mm. The electric field in this parallel plate waveguide is mainly TE mode (the electric field is normal to the plate). In this paper, all simulations are done in 3D. The refraction index of Maxwell's fish eye lens is [1]:

$$n(r) = \frac{2n_0}{1+(r/R_0)^2} \qquad (1)$$

Here $n_0$ is the refraction index on reference sphere and $R_0$ is the radius of the reference sphere. In our simulation, we set $R_0$ equal to the radius of PEC boundary $R_0$=R=5cm. In microwave region, we use PEC model for the metal. The source is an active coaxial cable with a core diameter of 0.5mm and an outer diameter of 1.68mm (a teflon isolator with relative electric permittivity 2.55 is filled between the PEC boundaries). The drain is also a coaxial cable with the same parameters as the source, but completely passive. The source and the passive drain are both inserted through the bottom PEC plate, exposing the cores into the waveguide structure by 4.5mm. The wavelength of the electromagnetic field produced by the source is 0.03m.

First we set an object (active coaxial cable) at (-0.03m, 0m) and a passive drain exactly at the image (0.03m, 0m). Furthermore we set one or more additional drains around the image to see what influence they have on the image resolution (see Fig. 1). If we set another additional passive drain at (0.03m, 0.0035m) (0.17λ apart from the image point and normal to the line connecting the image and the object), it has a very slight influence on the image resolution (see Fig. 1(a) and (e); as indicated in [10]). However, if we set the additional passive drain at (0.0265m, 0) (0.17λ apart from the image point and along the line connecting the object and the image), the influence becomes more obvious (see Fig. 1(b) and (f)). We notice that an additional drain located along the line connecting the object and the image degrades the image resolution more severely. This is due to the fact that the electromagnetic wave propagating directly from the the object to the image carries more energy. Furthermore, we also found that if four additional drains are set at (0.03m, 0.0035m), (0.03m, -0.0035m), (0.0265m, 0) and (0.0335m, 0) the image resolution will be degraded more severely than in the case when only one additional drain is introduced (see Fig. 1(c), (g) and (h)). If we set a drain array containing 9 passive drains around the image region and one drain is exactly at the image point, we found that multiple images which resemble the distribution of the passive drains will be formed (see Fig. 1(d)).

Now we set more than one passive drain around (but not exactly at) the image point. For this case our FEM simulation results show that one source will give multiple images (see Fig. 2). To verify this, we use a different numerical method, namely, the FDTD method, for a case when an array of passive drains is set around the image region of one object. The object is at (-0.03m, 0) while none of the passive drains are located exactly at the corresponding image point (0.03m, 0). The distance between two neighboring passive drains in the array is $0.1\lambda_0$. The simulation result is shown in Fig. 3, from which one sees that one object will give multiple images.

We have shown that we cannot use a drain array to detect the images in Maxwell's fish eye. Now we will study when the object is composed of various sources, whether we can use a single drain which is moved around the image region so as to scan the image points. For example, we set four source points at (-0.035m, 0), (-0.025m, 0), (-0.03m, 0.005m) and (-0.03m, -0.005m) in a

Maxwell's fish eye lens bounded in two PEC plates. The sources and the drain are coaxial cables as in our former simulations. As we can see from the simulation results in Fig. 4: if the drain is exactly located at one image point, it will form a sharp local field with high intensity around the drain (see Fig. 4(a); the field intensity is enhanced a bit at the source point due to the refocusing of the large field from the drain). When the drain is not located at any of the image points, it can still form a sharp local field with high intensity [see Fig. 4(b) (when the drain is in the middle of two sources) and (c) (the drain position is between the cases of (a) and (b))]. Therefore, if the object is composed of various sources, it is still not practical to use a single drain to scan the image points in a Maxwell's fish eye lens.

In a practical situation of imaging (e.g., bio-imaging), we do not know exactly where this object (source) is located, so the location of the corresponding image is also unknown. When one artificially sets many passive drains in the fish eye lens and then scans around the imaging region, one cannot obtain the field distribution of the images corresponding to the object points, but will obtain an imaging pattern of the artificial passive drains instead (see Fig. 2 and 3). Even if one of the drains is set exactly at one image point, the other additional drains will degrade the resolution (see Fig. 1(c) and (d)). Thus, it is not practical for imaging to introduce many passive drains in a Maxwell's fish eye lens, though they may sometimes sharpen the spot size of the image.

### 3. The role of the passive coaxial cable and Maxwell's fish eye medium

In our previous work, we have found that Maxwell's fish eye medium itself (without introducing drains) cannot give a subwavelength imaging as high order modes cannot reach the image point and contribute to the fine resolution [9]. Thus the sharp focusing spot of subwavelength size in the recent experiment [10] could be mainly due to the passive drains.

We also use a PEC pole to replace the passive coaxial cable in a Maxwell's fish eye lens bounded in two PEC plates. If we set a passive coaxial cable exactly at the image point (see Fig. 5(b)), the spot size around the image is FWHM=$0.0433\lambda_0$=$0.0637\lambda$. If we replace it with a PEC pole of identical size as the core inserted into the structure and at the same location (see Fig. 5(c)), the spot size around the image is nearly the same: FWHM=$0.0400\lambda_0$=$0.0588\lambda$.

Compared with our simulation in the previous section, we replace the fish eye medium in the plate waveguide structure with air. The source is still an active coaxial cable. If we set a passive coaxial cable with the same size as the source in this structure (i.e., air bounded by PEC), a sharp spot of subwavelength size (FWHM=$0.0500\lambda_0$=$0.0735\lambda$) can still be formed around the passive drain (see Fig. 6(a) and (c)). This shows that the subwavelength spot in Maxwell's fish eye is due to the drain but not to the fish eye medium. We also found that if we replace the passive coaxial cable in the "air lens" with a PEC pole whose dimensions are the same as the inner core of the source coaxial cable exposed into the waveguide structure, we can still get an equally sharp (if not sharper) subwavelength spot (FWHM=$0.0467\lambda_0$= $0.0687\lambda$) around the PEC pole in this "air lens" (see Fig. 6(b) and (d)).

Since the size of the PEC pole or the part of the coaxial cable exposed into the waveguide structure is much smaller than the wavelength, we can use a quasi-static model to describe the local field around the drain. The local field enhancement around a small PEC pole or a passive coaxial cable is easy to understand, just like that around the tip of a conical conductor [11]. We should note that 3D implementation of the fish eye lens is different from the 2D fish eye. For example, the 3D implementation may generate "resonances" between the drain's tip and the PEC outer "casing" (two PEC plates)--these would not be observed in the 2D original fish-eye lens. To see whether the "resonances" between the drain's tip and the PEC outer casing would also contribute to the crucial field enhancement, we plot the absolute value of electric field distribution around the passive coaxial cable used as a drain in the vertical direction (normal to the plane of the fish-eye lens). We found that the field enhancement is mainly localized around the metal tip (see Fig. 5(d)). Thus, for the high field intensity we can exclude the possibility of the tip's interaction with the PEC casing. Both a PEC pole and a coaxial cable can help to obtain a subwavelength spot size at the image in Maxwell's fish eye lens. If we apply an operator of $(\Delta_{x,y}+n^2k_0^2)$ upon field component $E_z$ in the cross section near the top PEC plate in the fish eye waveguide structure, we will obtain two approximate delta functions (not ideal delta functions) around the source and the image regardless of what drain we use (an coaxial cable or a PEC pole). Thus, both a coaxial cable and a PEC pole in the 3D waveguide structure can approximately mimic the drain in Leonhardt's 2D analytical solution for Maxwell's fish eye [4]. The phase difference between the source and the drain in the 2D analytical solution is $v\pi$ ($v=0.5(\sqrt{4k^2+1}-1)$), while in 3D plate waveguide structure it may not be $v\pi$ (depending on the specific structure; about 0 for a PEC pole used as a drain and $\pi/6$ for a coaxial cable used as a drain in the structure of Fig. 5). The phase delay $v\pi$ between the source and the drain is due to the light propagation in the index profile (1) in 2D space. However, in 3D space the light no longer propagates in one plane, and consequently the phase difference is no longer $v\pi$ but related to the specific 3D structure.

A coaxial cable and a PEC pole in 3D waveguide structure can both approximately mimic the drain. However they have a main difference: a coaxial cable can extract the wave (some power will leave the waveguide structure through it), while a PEC pole cannot extract any wave. Thus the subwavelength focusing achieved in this waveguide model is due to some local field enhancement at the image (e.g., by using a metal tip) instead of some outlet of the wave.

It should be noted that if we replace the coaxial cable at the image with a PEC pole, after a stationary configuration has been formed, the coaxial cable at the object radiates energy (0.0149W) in the first half period of the power flux and absorbs energy (-0.0164W) in the second half period of the power flux (the period of the power flux is half of the harmonic oscillation period of the source) while the passive PEC pole does not absorb any energy. The coaxial cable at the object can be either an inlet or an outlet of the wave. This has also been verified by our FDTD simulation (see Fig. 7).

## 4. The finite resolution of Maxwell's fish eye lens with passive drains

The resolution is the ability of an optical system to distinguish the smallest distance of two close objects [3]. Perfect imaging requires infinite resolution, which means that the smallest distance this optical system can distinguish has to be infinitesimal. In our previous work [9] we have shown that Maxwell's fish eye's resolution is limited if there is no drain at the image. In this section, we analyze if Maxwell's fish eye can make a perfect imaging when some drains are set exactly at the image points. We use Leonhardt's 2D analytical solution [4] in Maxwell's fish eye lens to obtain the field distribution, when we set two close objects and two drains at the image points.

When we set a line current source at $z_0$ and a drain at $-z_0$, the field distribution in 2D Maxwell's fish eye bounded by a PEC boundary with radius $R=R_0$ can be written as [4]:

$$E(z) = \frac{1}{4\sin(v\pi)} \{P_v(\zeta(w(z))) - P_v(\zeta(w(1/z^*))) + e^{iv\pi}P_v(-\zeta(w(1/z^*))) - e^{iv\pi}P_v(-\zeta(w(z)))\} \quad (2)$$

where $P_v(\zeta)$ is the Legendre functions with $v=0.5(\sqrt{4k^2+1}-1)$, $w(z)$ means Möbius transformation $w(z)=(z-z_0)/(z_0^*z+1)$ with $z=x+iy$, and $\zeta(w)=(|w|^2-1)/(|w|^2+1)$. We should note that the unit of the length is $R_0$ in all the above formulas.

For numerical calculation, we set two source points at (-0.03m, 0) and (-0.03m-$\Delta$, 0) and two drains at (0.03m, 0) and (0.03m+$\Delta$, 0) in 2D Maxwell's fish eye with PEC boundary R=$R_0$=5cm. $\Delta$ is the distance between the two objects (the same as the distance between the two images). The wavelength of the source radiation is 3cm. From analytical solution (2) we obtain the total field distribution and plot along x direction for different $\Delta$ in Fig. 8. From this figure we can see that it cannot distinguish two image points if $\Delta$ is very small (e.g., less than 0.01$\lambda_0$) regardless if the sources are coherent or not. This indicates that even if we set all passive drains at the exact right image points, the resolution of Maxwell's fish eye is limited. Actually it is not meaningful to talk about the resolution after knowing that the physical reason for these subwavelength spots are merely local field enhancement around the passive drains.

## 5. Conclusion

In this paper we have shown that if some passive drains are set at the exact image point and some additional nearby points, the resolution of an object will be degraded by those additional passive drains. We have also found that the drains located along the line connecting the object point and the image point degrade the image resolution more severely. If many passive drains are set around the image point of one object, one would observe multiple images which resemble the distribution of the passive drains. Thus it is not practical in an imaging application to introduce many passive drains in Maxwell's fish eye lens.

We have also shown that the subwavelength resolution achieved in Maxwell's fish eye lens [4, 10] is due to the drain (conjugating with the object) but not the fish eye medium [9]. If we keep all the setup in the recent microwave experiment [10] unchanged except replacing the fish eye medium with air, we can still obtain a subwavelength spot around the passive drain. We have found that a subwavelength spot at the image point can still be formed if the coaxial cable served as a passive drain at the image point is replaced by a PEC pole (identical to the core of the coaxial cable inserted into the waveguide structure filled with fish eye medium or air). A PEC pole can do equally well as a coaxial cable for achieving a sharp field spot though a PEC pole cannot serve as an outlet of the electromagnetic wave.

We have found that the subwavelength field spot in the recent experiments [10] is neither due to the fish eye medium itself nor the ability of the outlet in extracting wave, but due to the local field enhancement at the metal pole. At an optical frequency, similar localized field enhancement could occur at a gold (plasmonic) nano particle set at the image position if the wavelength of the source light is at the resonance wavelength of the localized surface plasmon polarition (SPP) mode of the gold nano particle. The focusing effect in [12] is not of subwavelength (the spot size is about 2 μm for the wavelength of 1.55 μm) and this may be because the wavelength is not at the resonance wavelength of the localized SPP mode of the gold nano particle.

**Acknowledgments**

The work is partly supported by the National Basic Research Program, the National Natural Science Foundations of China, and the Swedish Research Council (VR) and AOARD.

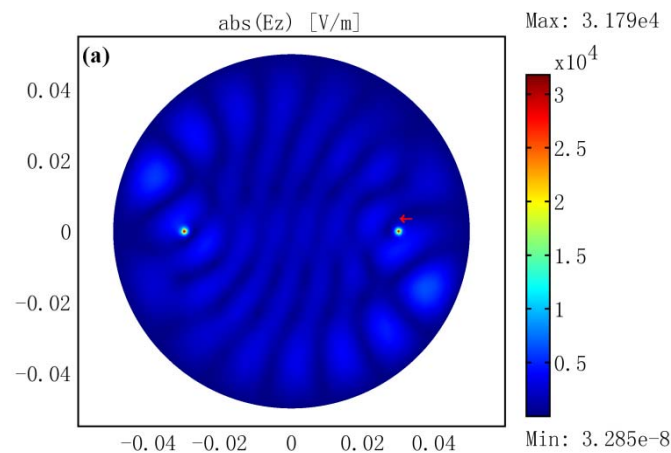

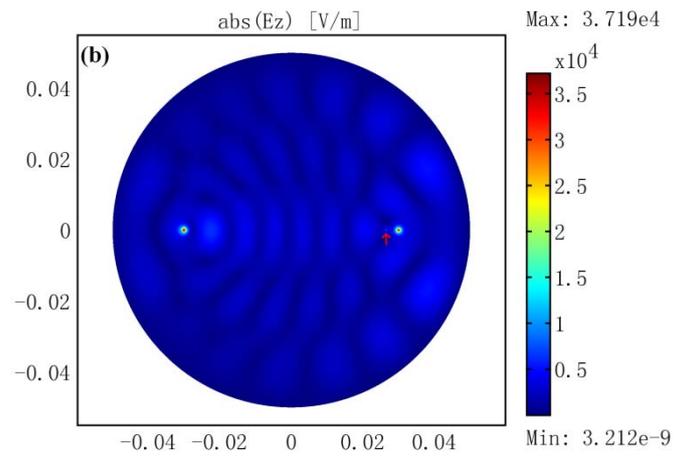

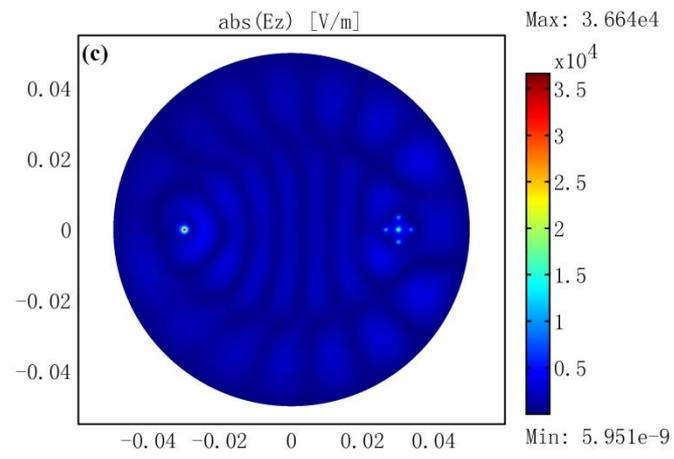

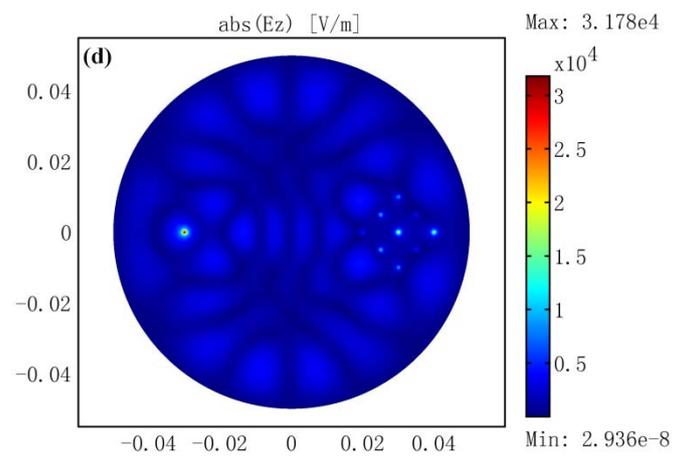

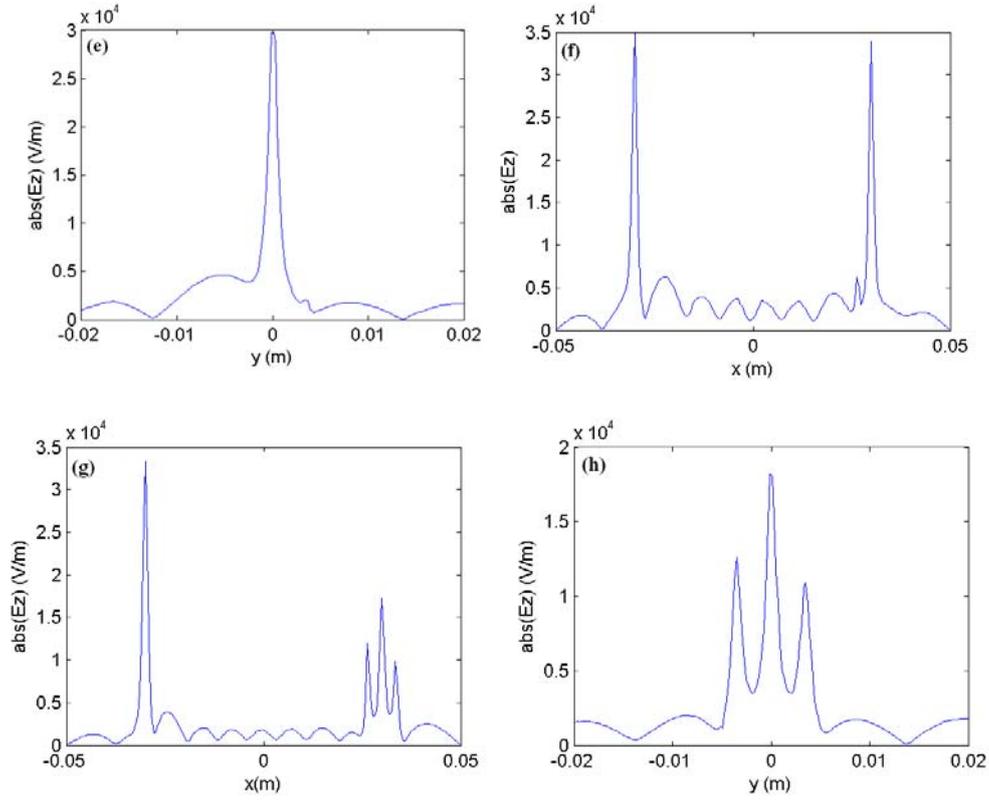

Fig. 1. FEM simulation results. The source and passive drains are inserted from the bottom PEC plate by 4.5mm. Both coaxial cables have the same parameters: the diameters are 0.5mm for the inner PEC and 1.68mm for the outer PEC. A teflon isolator with a relative electrical permittivity of 2.55 is filled between two PEC boundaries. The simulation wavelength is $\lambda_0$=0.03m, and the size of Maxwell's fish eye lens is R=1.667$\lambda_0$. The source is at (-0.03m, 0) and a drain is at (0.03m, 0). (a), (b), (c), (d) Distribution for the absolute value of the electric field on the cross section near the top PEC plate. (a) When we set another additional drain at (0.03m, 0.0035m). (The location of the drain is marked with a red arrow.) (b) When we set another additional drain at (0.0265m, 0). (The location of the drain is marked with a red arrow.) (c) When we set four additional drains at (0.03m, 0.0035m), (0.03m, -0.0035m), (0.0265m, 0) and (0.0335m, 0). (d) When we set a drain array of 9 passive coaxial cables with a neighboring distance of 0.2357$\lambda_0$. The central pixel of the array is at the image point (0.03m, 0). (e) The absolute value of electric field distribution along line x=0.03m of (a). (f) The absolute value of electric field distribution along line y=0 of (b). (g) and (h) The absolute value of the electric field distribution along line y=0 and x=0.03m of (c).

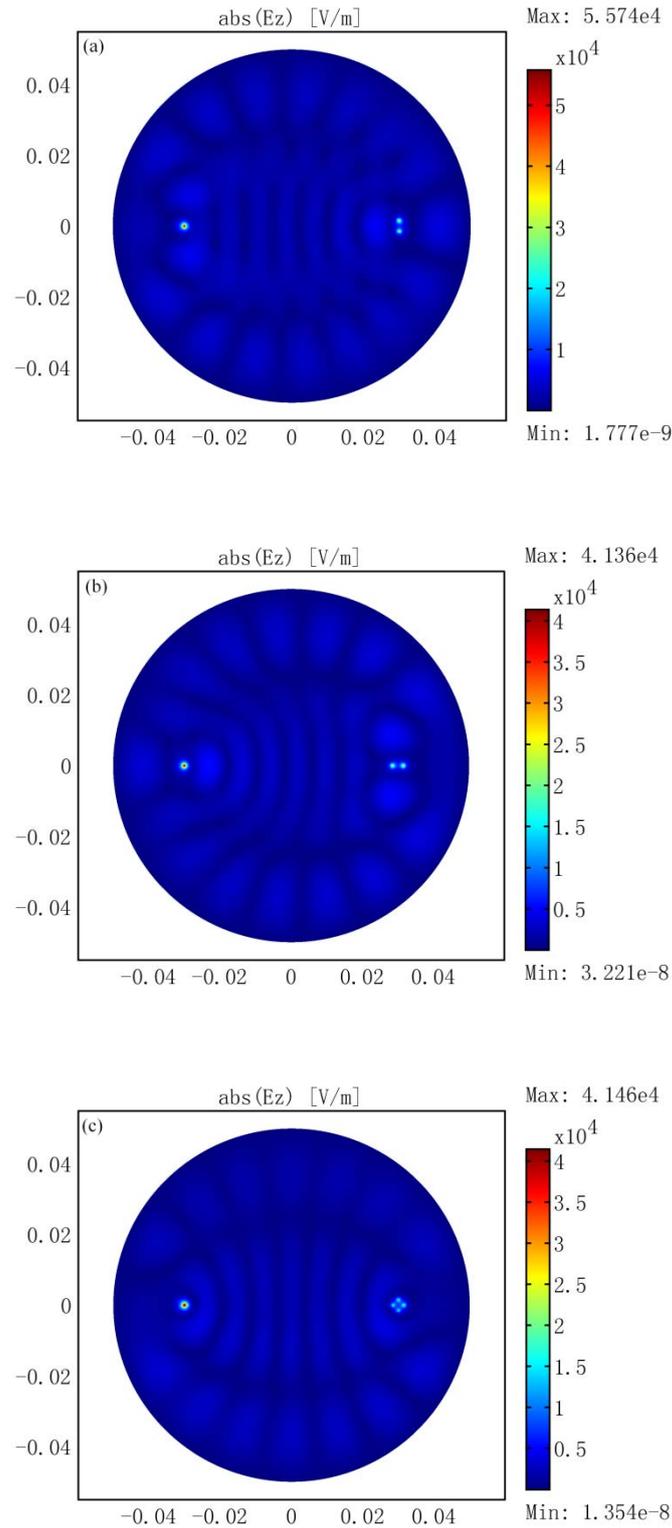

Fig. 2. FEM simulation results at the cross section near the top PEC plane for different drain situations. The absolute value of the electric field distribution in Maxwell's fish eye lens bounded between two PEC plates. The source and the drains are coaxial cables with the same parameters as used for Fig. 1. The simulation wavelength is $\lambda_0=0.03$m, and the size of Maxwell's fish eye lens is $R=1.667\lambda_0$. The source is at (-0.03m, 0). (a) When two drains are set at (0.03m, 0.0015m) and (0.03m, -0.0015m). (b) When two drains are set at (0.0315m, 0) and

(0.0285m, 0). (c) When four drains are set at (0.0315m, 0), (0.0285m, 0), (0.03m, 0.0015m) and (0.03m, -0.0015m).

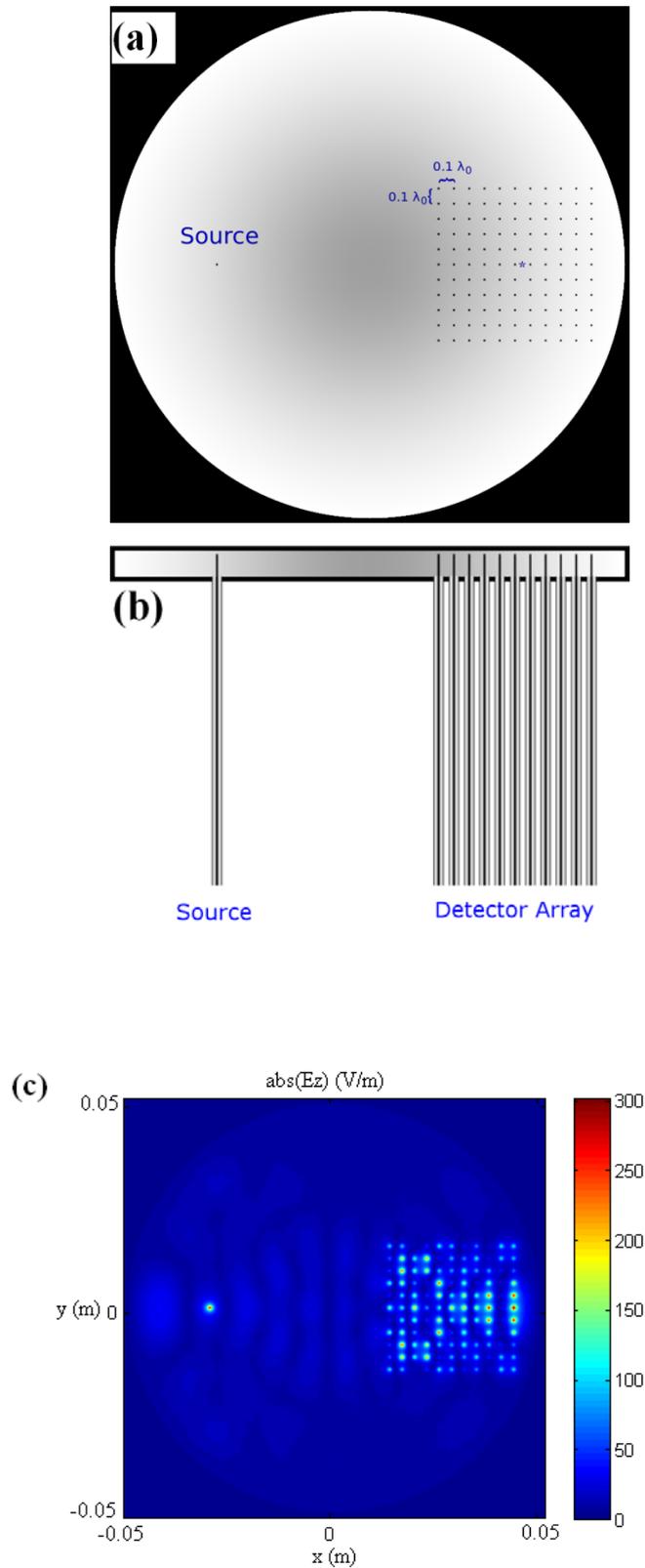

Fig. 3. FDTD simulation structure and results: (a) Cross section plot of the drain array in the fish eye lens. (b) Side view of the structure. (c) The cross section near the top PEC plane. The absolute value of the electric field (time

average value in one harmonic period) in Maxwell's fish eye lens bounded between two PEC plates. The source and the drains have the same parameters as used for Fig. 1. The wavelength is $\lambda_0=0.03$m, and the size of Maxwell's fish eye lens is $R=1.667\lambda_0$. The source is at (-0.03m, 0) and the distance between two close pixel in the drain array is $0.1\lambda_0$

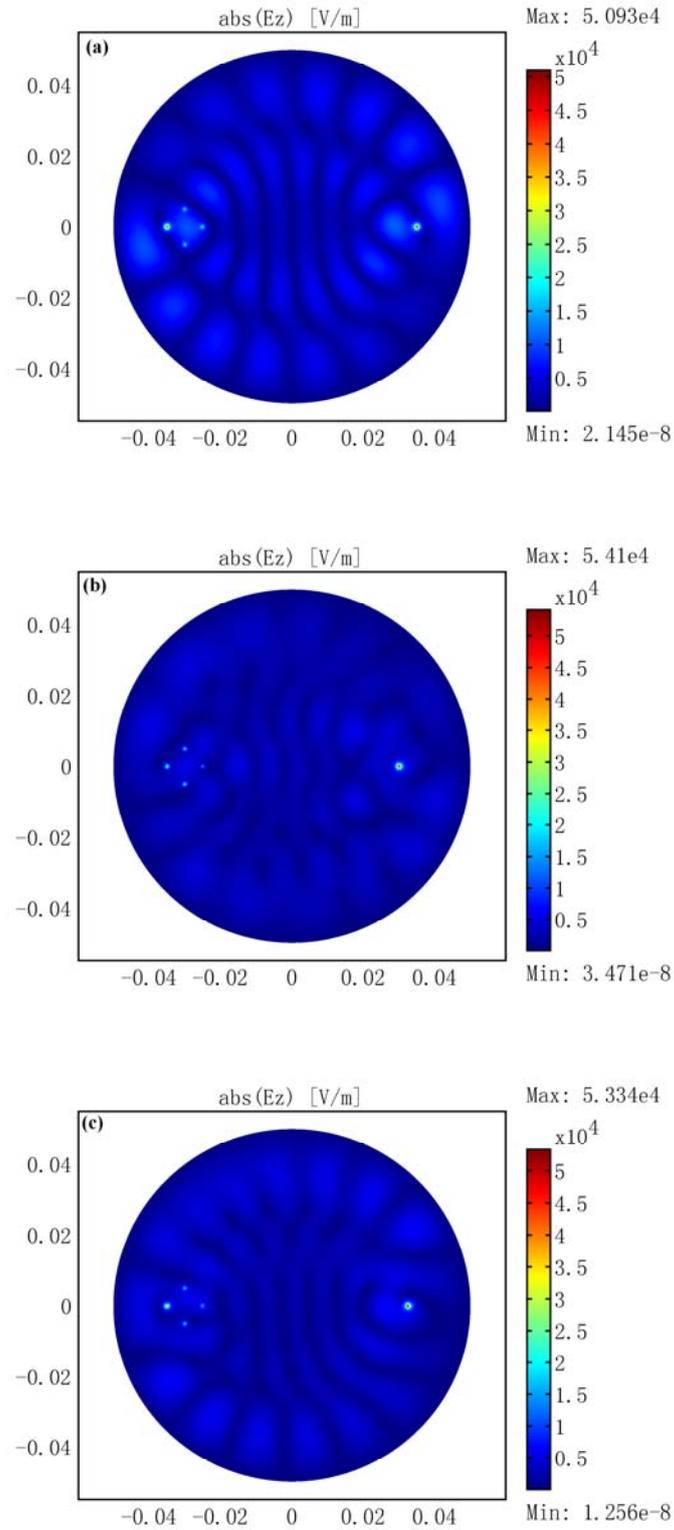

Fig. 4. FEM simulation results when the object is composed of various sources. We set a single drain localized at different positions around the image region. We set four active coaxial cables (sources) at (-0.035m, 0), (-0.025m, 0), (-0.03m, 0.005m) and (-0.03m, -0.005m). The single drain is also a passive coaxial cable that is moved around the image region. The parameter of the coaxial cables and the fish eye lens are the same as Fig.1. The location of the single drain is at (a) (0.035m, 0) (an image point), (b) (0.03m, 0) and (c) (0.0325m, 0).

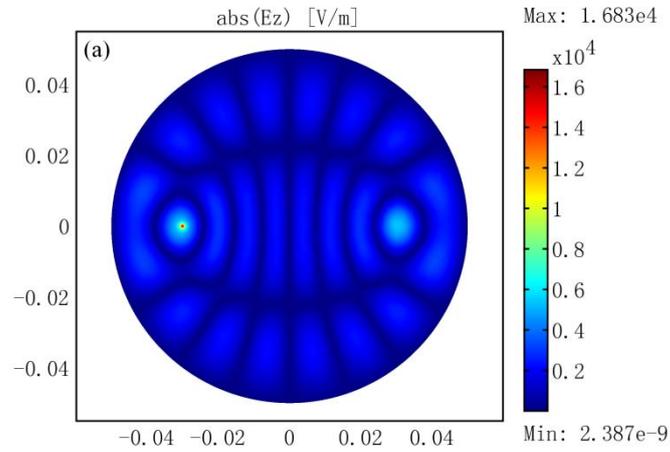

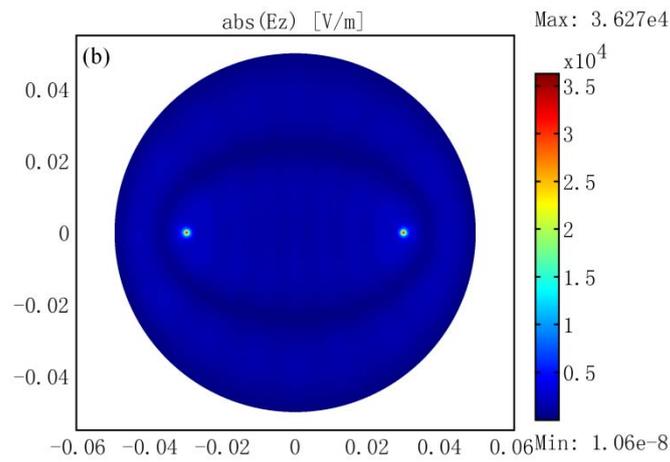

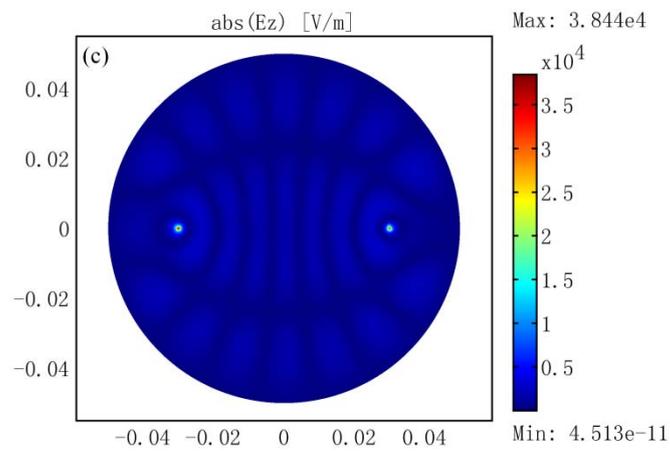

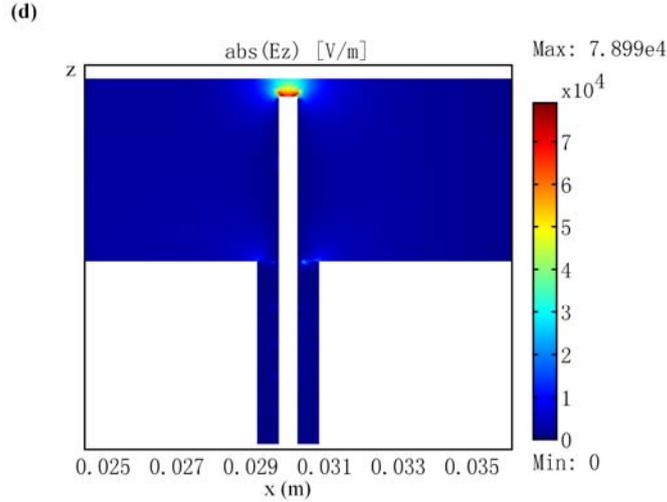

Fig. 5. FEM simulation result for the absolute value of electric field distribution on the cross section near the top PEC plane of the Maxwell's fish eye lens bounded between two PEC plates. The source is an active coaxial cable at (-0.03m, 0) with the same parameters as used for Fig. 1. The wavelength is $\lambda_0$=0.03m, and the size of Maxwell's fish eye lens is R=1.667$\lambda_0$. (a) Without any passive drain at the image point. The size of the image spot is FWHM= 0.3233$\lambda_0$= 0.4754$\lambda$. (b) A passive coaxial cable (with the same parameters of the source) at the image point (0.03m, 0). The size of the image spot is FWHM=0.0433$\lambda_0$=0.0637$\lambda$. (c) A cylindrical PEC pole is set at image point (0.03m, 0) with the same size of the inner wire of the source coaxial cable exposed into the waveguide structure: radius 0.25mm and height 4.5mm. The size of the image spot is FWHM=0.0400$\lambda_0$=0.0588$\lambda$. (d) The cross section (x-z plane) of the absolute value of the electric field around the passive coaxial cable at (0.03m, 0).

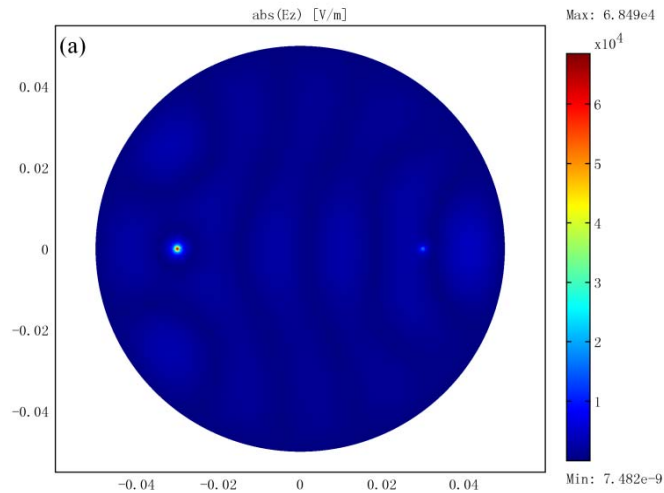

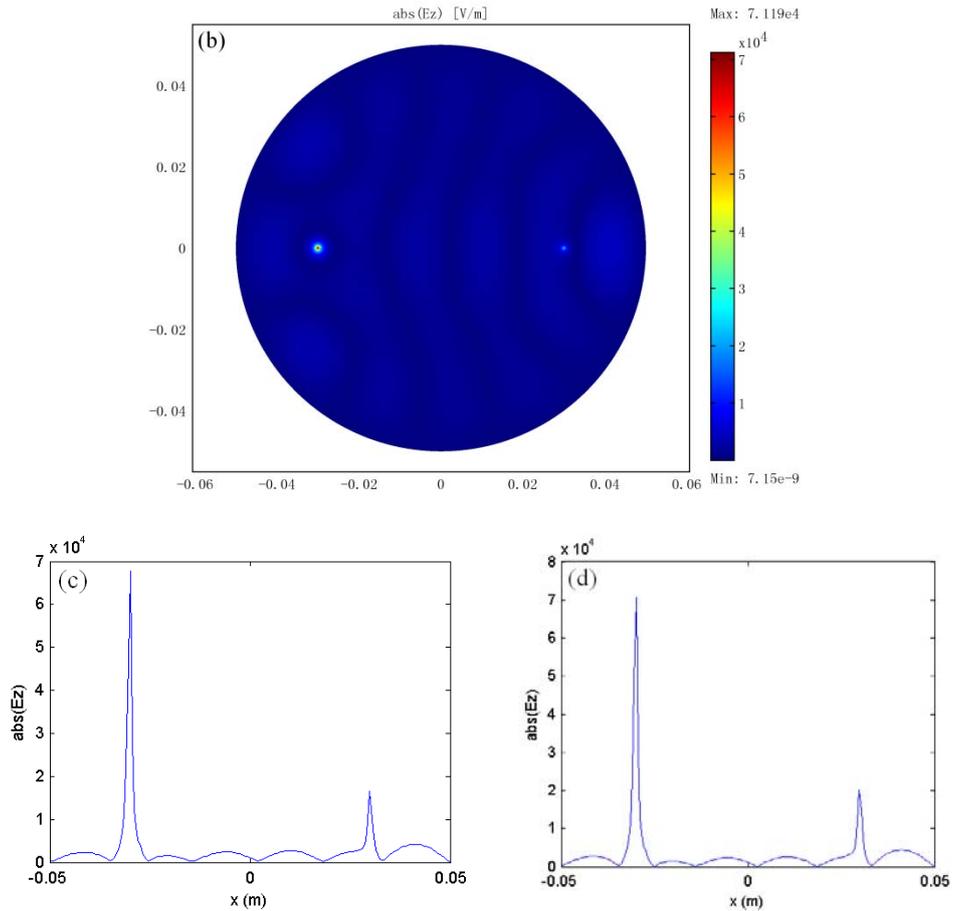

Fig. 6. FEM simulation result for the field distribution in the "air lens" (the Maxwell's fish eye medium is replaced with air). The cylindrical PEC boundary has a radius of 0.05m and a height of 5mm. The source at (-0.03m, 0) is still an active coaxial cable with the same parameters as used for Fig. 1. The wavelength is $\lambda_0$=0.03m. (a) and (c): We set a passive coaxial cable (with the same size as the source) at (0.03m, 0). (b) and (b): We set a PEC pole (with the same size as the inner wire of the source coaxial cable) exposed into the waveguide structure at (0.03m, 0). (a) and (b): The absolute value of electric field on a cross section near the top PEC plate. (c) and (d): The absolute value of electric field along x axis of (a) and (b), respectively.

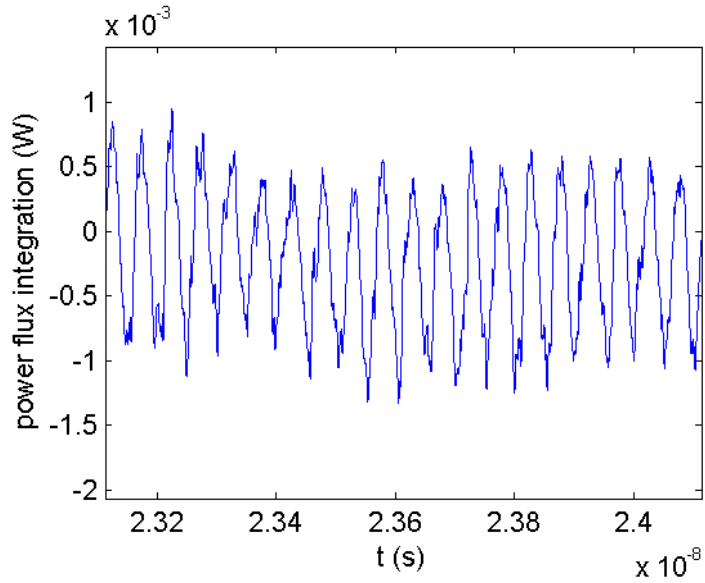

Fig. 7. FDTD simulation results after a very long time period (a steady state has been approximately formed). The time-varying integration of the power flux at the interface between the waveguide structure and the coaxial cable used as the object (source). The parameters for the waveguide structure and the coaxial cable are the same as those used for Fig. 1. The drain is a PEC pole (with the same size of the inner core of the source coaxial cable) exposed into the waveguide structure. The wavelength of the coaxial cable's radiation is $\lambda_0=0.03$m.

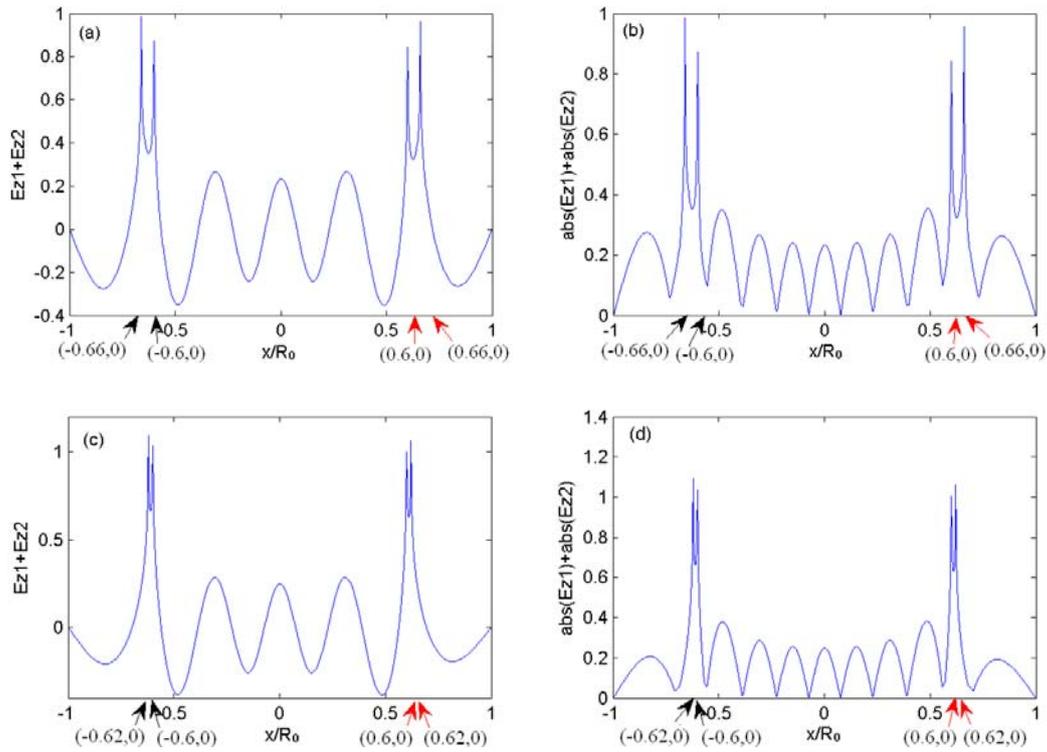

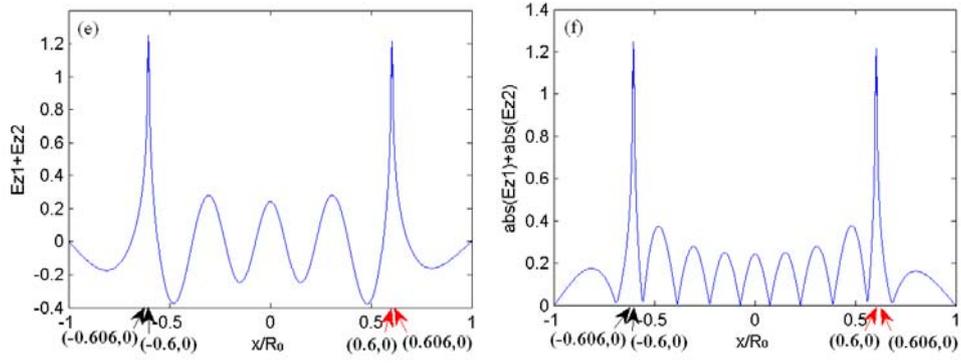

Fig. 8. Analytical results [based on Eq. (2)] for the field distribution along the x direction in 2D Maxwell's fish eye lens with PEC boundary $R=R_0=5$cm. The simulation wavelength is $\lambda_0=3$cm. We use black and red arrows to indicate the location of the objects and the images, respectively. We set two sources points at (-0.03m, 0) and (-0.03m-$\Delta$, 0) and two drains at (0.03m, 0) and (0.03m+$\Delta$, 0). In (a) and (b) we choose $\Delta=0.1\lambda_0$. Two separate peaks can obviously be seen around the images in this case. In (c) and (d) we choose $\Delta=0.033\lambda_0$. We can still see two separate peaks around the objects and images, however, it is hard to resolve them due to the fact that the overlapped energy of the two peaks is too large. In (e) and (f), we choose $\Delta=0.01\lambda_0$. The two peaks can not be distinguished because the distance is too small. (a), (c), (e) are for coherent light imaging: the distribution of total electric field $E_z$ along x direction. (b), (d), (f) are for incoherent light imaging: the absolute value of total electric field $E_z$ along x direction.